\documentclass[12pt]{article}

\usepackage{graphicx}
\usepackage{epsfig}
\usepackage[latin1]{inputenc}
\usepackage{xcolor}
\usepackage{units}
\usepackage{xspace}
\usepackage{rotating}
\usepackage{cleveref}

\usepackage{verbatim}

\def\xmax{\ifmmode {X_\mathrm{max}}\else{$X_\mathrm{max}$}\fi\xspace}%
\def\meanXmax{\ifmmode {\langle X_\mathrm{max}\rangle}\else{$\langle X_\mathrm{max}\rangle$}\fi\xspace}%
\def\eV{\ifmmode {\mathrm{\ e\kern -0.1em V}}\else\textrm{e\kern -0.1em V}\fi\xspace}%
\newcommand{\energyEV}[1]{\unit[$10^{#1}$]{\eV}}
\newcommand{\energy}[1]{$10^{#1}$}

\title{Cosmic rays: the spectrum and chemical composition from $10^{10}$ to $10^{20}$ eV}

\author{C. J. Todero Peixoto$^{\ddagger,\star}$, Vitor de Souza$^{\ddagger}$ \\ and Peter L. Biermann$^\dagger$\thanks{Also at Physics Dept., K.I.T. Karlsruhe, Germany; Dept. Physics \& Astronomy, Univ. Alabama, Tuscaloosa, AL, USA; Dept. Physics \& Astronomy, Univ. Bonn, Germany}\\
{\footnotesize\textit{$^{\ddagger}$Universidade de S\~ao Paulo, Instituto de F\'isica de S\~ao Carlos, Brazil}}\\
{\footnotesize\textit{$^{\star}$Universidade de S\~ao Paulo, Depto. de Ci\^encias B\'asicas e Ambientais}}\\
{\footnotesize\textit{Escola de Engenharia de Lorena, Brazil}}\\
{\footnotesize\textit{$^\dagger$Max-Planck-Institute for Radioastronomy, Auf dem H\"ugel 69, 53121 Bonn, Germany}}\\
}

\date{\today}

\begin{document}
\maketitle

\begin{abstract}
The production of energetic particles in the universe remains one of the great mysteries of modern science. The mechanisms of acceleration in astrophysical sources and the details about the propagation through the galactic and extragalactic media are still to be defined. In recent years, the cosmic ray flux has been measured with high precision in the energy range from \energy{10} to \energyEV{20.5} by several experiments using different techniques. In some energy ranges, it has been possible to determine the flux of individual elements (hydrogen to iron nuclei). This paper explores an astrophysical scenario in which only our Galaxy and the radio galaxy Cen A produce all particles measured on Earth in the energy range from \energy{10} to \energyEV{20.5}. Data from AMS-02, CREAM, KASCADE, KASCADE-Grande and the Pierre Auger Observatories are considered. The model developed here is able to describe the total and individual particle flux of all experiments considered. It is shown that the theory used here is able to describe the smooth transition from space-based to ground-based measurements. The flux of each element as determined by KASCADE and KASCADE-Grande and the mass sensitivity parameter \xmax measured by the Pierre Auger Observatory above \energyEV{18} are also explored within the framework of the model. The transition from \energy{16} to \energyEV{18} is carefully analyzed. It is shown that the data measured in this energy range suggest the existence of an extra component of cosmic rays yet to be understood.
\end{abstract}

{\footnotesize Keywords: cosmic rays, galaxies: active - galaxies, galaxies: jets - galaxies: starburst}

\section{Introduction}

The currently accepted view is that the cosmic rays are produced in active astrophysical objects: supernovae, gamma-ray bursts, active stars in binary systems, pulsars, active galactic nuclei, quasars, radio galaxies, and large-scale structure shocks. Possible sources of cosmic rays in the Galaxy include supernova explosions, pulsars and the Galactic nucleus, which contains a super-massive black hole. Traditional models of stochastic acceleration assume the interaction of particles with magnetic fields, according to the Fermi mechanisms~\cite{Fermi1949}, where the particles would be accelerated in collisions with magnetic moving clouds. In principle, this same mechanism, with few changes, can accelerate particles in shock waves in supernovae, gamma-ray bursts, Wolf-Rayet star winds, active galactic nuclei, radio galaxies and other sites. This paper explores a model originally proposed in reference~\cite{Stanev_Biermann_Gaisser_1993} in which a combination of four main components is used to explain the cosmic ray spectrum. Figure~\ref{fig:theory:original} shows a schematic picture of the main features of the model. The details of the model are discussed in the next section. It has been previously shown in reference~\cite{Biermann_Vitor_2012} that this model is able to describe the total flux of cosmic rays up to \energyEV{20} using one extra parameter: an energy shift factor of 2800 caused by re-acceleration of galactic seed particles in the jets.

The calculations presented here extend the validity of the model in two ways. Firstly, an analysis is presented of the cosmic ray flux with energies between \energy{10} and \energyEV{15}. The recently published data from AMS-02~\cite{AMS-02_ICRC_2013} and CREAM~\cite{CREAM_2041-8205-714-1-L89, CREAM_0004-637X-707-1-593} are used. These experiments are able to discriminate with high precision the individual elements of the cosmic ray composition, measuring the flux of each particle type. Given the high statistics achieved by these space and balloon-borne experiments, the data constrain severely the contribution of each particle to the total flux. It is shown here for the first time how this model describes very well the energy range from \energy{10} to \energyEV{15}, which includes the transition regime of space and balloon experiments to ground-based observatories. This first analysis represents an extension in the energy range for which this model is able to describe the measured total flux of cosmic rays.

In a second analysis, we show the agreement of the model concerning the flux of each element: hydrogen to iron nuclei. For energies below \energyEV{15} the analysis can be done taking into account the individual flux measured by AMS and CREAM for each element. For energies above \energyEV{15} the analysis is done through considering indirect composition measurements. The data from KASCADE and KASCADE-Grande experiments are compared to the model predictions fixed by the space and balloon experiments. It is shown that a continuation extension of the AMS and CREAM (\energy{10} to \energyEV{14}) to KASCADE and KASCADE-Grande (\energy{15} to \energyEV{18}) data is very hard to achieve due to the high flux of heavy elements measured by KASCADE-Grande in the energy range from \energy{17} to \energyEV{18}. The model prediction is also compared to the evolution of the mean depth of the shower maximum (\xmax) measured by the Pierre Auger Observatory~\cite{bib:auger:nim}.

Section~\ref{sec:method} reviews the original model and its tests. Section~\ref{sec:spectra} shows the first analysis in which the model is extended to low energies and section~\ref{sec:mass} shows the analysis regarding the cosmic ray composition. Section~\ref{sec:conclusion} summarizes the main conclusions of the paper.

\section{The original model}
\label{sec:method}

\begin{figure}
\centerline{\includegraphics[width=13cm]{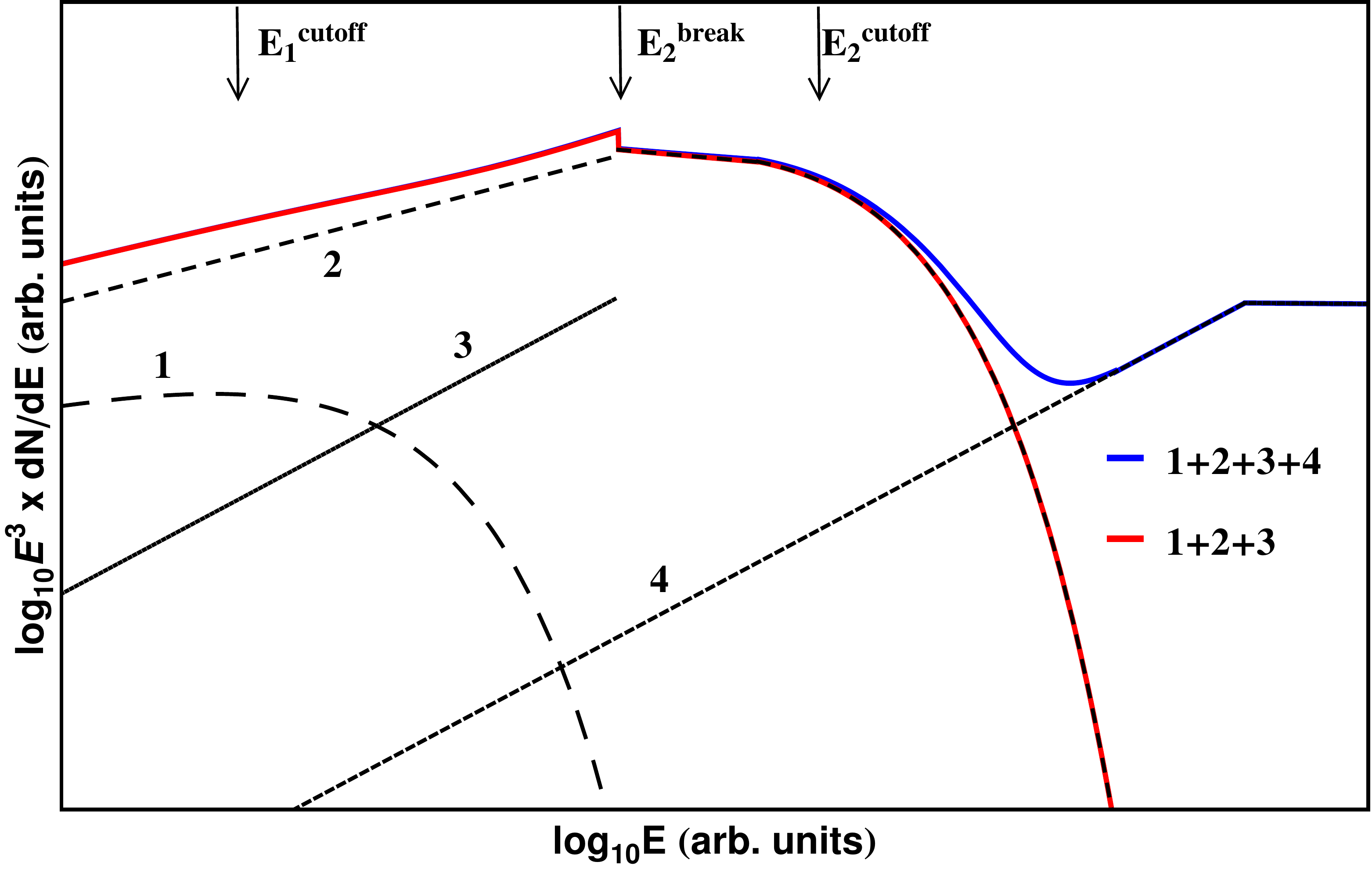}}
\caption{Schematic representation of the energy spectrum as predicted by the original model~\cite{Stanev_Biermann_Gaisser_1993}. Four curves are  due to the four components: 1) supernova explosions in the interstellar medium (Sedov phase), 2) supernova explosion into the stellar winds (Wolf-Rayet), 3) polar cap component of supernova explosion and 4) extragalactic contribution. $E_1^{cutoff}$ is the cutoff energy for component 1, $E_2^{break}$ marks a step in the flux of component 2 due to a regime transition in the drift gain, and $E_2^{cutoff}$ is the cutoff energy for component 2. Figure adapted from reference \cite{Stanev_Biermann_Gaisser_1993}.}
\label{fig:theory:original}
\end{figure}

Figure~\ref{fig:theory:original} shows the main features of the original model which was proposed to explain the observed features of cosmic rays in the energy range from $10^{14}$ to $10^{18}$~eV. The model is based on three (1, 2 and 3) galactic and one (4) extragalactic component. The phenomena contributing to the acceleration of particles are: a) supernova explosions into the interstellar medium, b) supernova explosions into the stellar wind, and c) powerful jets of radio-galaxies. In summary, supernova explosions generate the galactic cosmic rays up to $10^{17-18}$ eV and radio-galaxies jets re-accelerate galactic cosmic rays to the highest energies $10^{18-20}$ eV. In this scenario, the main source of extragalactic cosmic rays is Cen A.

Label 1 in figure~\ref{fig:theory:original} is the resulting energy spectrum outcome of supernova explosions into the interstellar medium. The maximum energy that a cosmic ray can be accelerated in a supernova shock taking into account the Sedov expansion into the interstellar medium was calculated in reference~\cite{Lagage_Cesarsky_1983}. The produced spectrum has index value proposed to be around $-2.75$ and an exponential cutoff. This component can be written as:

\begin{equation}
(dN/dE)_1 = A_1 \cdot E^{-2.75} \cdot \exp{-E/E_1^{cutoff}},
\end{equation}
where $A_1$ is the normalization of the flux and $E_1^{cutoff}$ is the cutoff energy. The cutoff energy is predicted to be proportional to charge ($E_1^{cutoff} \propto Z$).

Label 2 in figure~\ref{fig:theory:original} is the resulting energy spectrum outcome of supernova explosions into the stellar wind, like a Wolf-Rayet star explosion. The produced spectrum has an index around $-2.67$ for energies smaller than $E_2^{break}$ and index around $-3.07$ for energies smaller than $E_2^{cutoff}$ which determines the exponential cutoff of the flux. $E_2^{break}$ and $E_2^{cutoff}$ are both predicted to be proportional to charge. The existence of two regimes is due to the dependence of the acceleration efficiency to the particle drift gain~\cite{Stanev_Biermann_Gaisser_1993}. This component can be written as:

\begin{equation}
  (dN/dE)_2 =  \left\{ \begin{array}{ll}
     A_2 \cdot E^{-2.67}  & \mbox{ if $E < E_2^{break}$} \\
     B_2 \cdot E^{-3.07} \cdot \exp{ -E/E_2^{cutoff} } & \mbox{ if $E > E_2^{break}$}
     \end{array} \right.
\end{equation}

where $A_2$ and $B_2$ are normalization of the flux.

Label 3 in figure~\ref{fig:theory:original} is an extra component resulting from the outcome of a supernova explosions into the stellar wind. In the final stage of the very massive stars there is a connection between rotation and magnetic field. This magneto-rotational mechanism for massive stars explosions was first proposed by Bisnovatyi-Kogan~\cite{Bisnovatyi-Kogan-1970,Bisnovatyi-Kogan-2007} and seems consistent with the energy/charge ratio for the heavy elements~\cite{PhysRevLett.103.061101,2041-8205-710-1-L53,0004-637X-725-1-184,Gupta-Nath-Biermann-2013}. This connection produces a polar cap component relevant in the region where the radial field $B_r\sim 1/r^2$ dominates. The energy spectrum index was predicted to be around $-2.33$ with a sharp cutoff at $E_3^{cutoff}$:

\begin{equation}
(dN/dE)_3 = \begin{array}{ll} A_3 \cdot E^{-2.33} & \mbox{ if $E < E_3^{cutoff}$} \end{array}
\end{equation}
where $A_3$ is the normalization of the flux and $E_3^{cutoff}$ is the cutoff energy. The original model predicts  $E_2^{break} = E_3^{cutoff}$.

This model proposal has used the concept, that transport of cosmic rays is governed by Kolmogorov turbulence, and that the secondary particles are produced in interactions near the source~\cite{PhysRevLett.103.061101,bib:b:2001}.

Label 4 in figure~\ref{fig:theory:original} is an extragalactic component proposed to explain the highest energy range. Radio Galaxies such as the Fanaroff-Riley class II have hot spots at the end of linear radio features, which are considered to be highly collimated plasma jets. The evolution of these powerful radio galaxies can explain the spectrum to energies above \energyEV{18}~\cite{Rachen_Biermann_1993_1,Rachen_Biermann_1993_2}. The predicted index of the generated energy spectrum is approximately $-2$.

It has been shown that the same mechanism rescaled in energy by a factor of 2800 can accelerate particles up to \energyEV{20}~\cite{Biermann_Vitor_2012}. The argument was based on the re-acceleration of the original galactic seeds in the jets of radio galaxies. Interpretation of observations to derive the central Lorentz factor required in the relativistic jets emanating from near super-massive black holes in Active Galactic Nuclei (AGN) suggest values of up to $\gamma_j \, = \, 100$~\cite{bib:gopal:2014,bib:kellermann}. As Gallant \& Achterberg~\cite{MNR:MNR2566} as well as~\cite{MNR:MNR4851} have shown, the acceleration of particles in relativistic shocks, clearly possible in AGN jets up to maximally the Lorentz factor of the jet itself, gives an increase in energy/momentum by $\gamma_j^2$ in a single first step, and for all subsequent steps considerably less.  So we use here what could be called the ``single kick approximation", namely only that single first step.  Observations suggest that jets are energized intermittently (see, e.g., the radio galaxy Her A,~\cite{MNR:MNR6469}).  Such extreme Lorentz factors may be possible in the ``working surface" of a freshly energized jet.

\section{Comparison of the model to measured energy spectra of elements}
\label{sec:spectra}

In previous publications, the predictions of the original model~\cite{Stanev_Biermann_Gaisser_1993} and the predictions of its extrapolation to the highest energies~\cite{Biermann_Vitor_2012} were compared to the total flux of cosmic ray particles measured by several experiments. These comparisons have been able to show the general validity of the model. Nevertheless they have not been able to remove the intrinsic degeneracy of the model concerning the abundance of each element. If only the sum of all elements is verified, several predictions of the model, for instance, rigidity dependencies cannot be tested. Besides that, the large number of free parameters to fit the total flux reduces the significance of the final results. In this paper, the tests are done using the most up-to-date spectrum of each element as measured by several experiments. The intention of the analysis presented in this section is to remove the freedom in describing the total flux as presented in the previous studies.

The original model flux is dominated by component 2 see figure~\ref{fig:theory:original}. Component 3 causes a break in spectrum which could be used to discriminate the model explored here from other traditional models, i.e., Peters cycles~\cite{bib:peter:cycles,bib:stanev:peter:cycles}. The step in the energy spectrum caused by components 1 and 2 requires a very precise measurement of the energy spectrum of each element in order to be tested.

\begin{figure}
\centerline{\includegraphics[width=13cm]{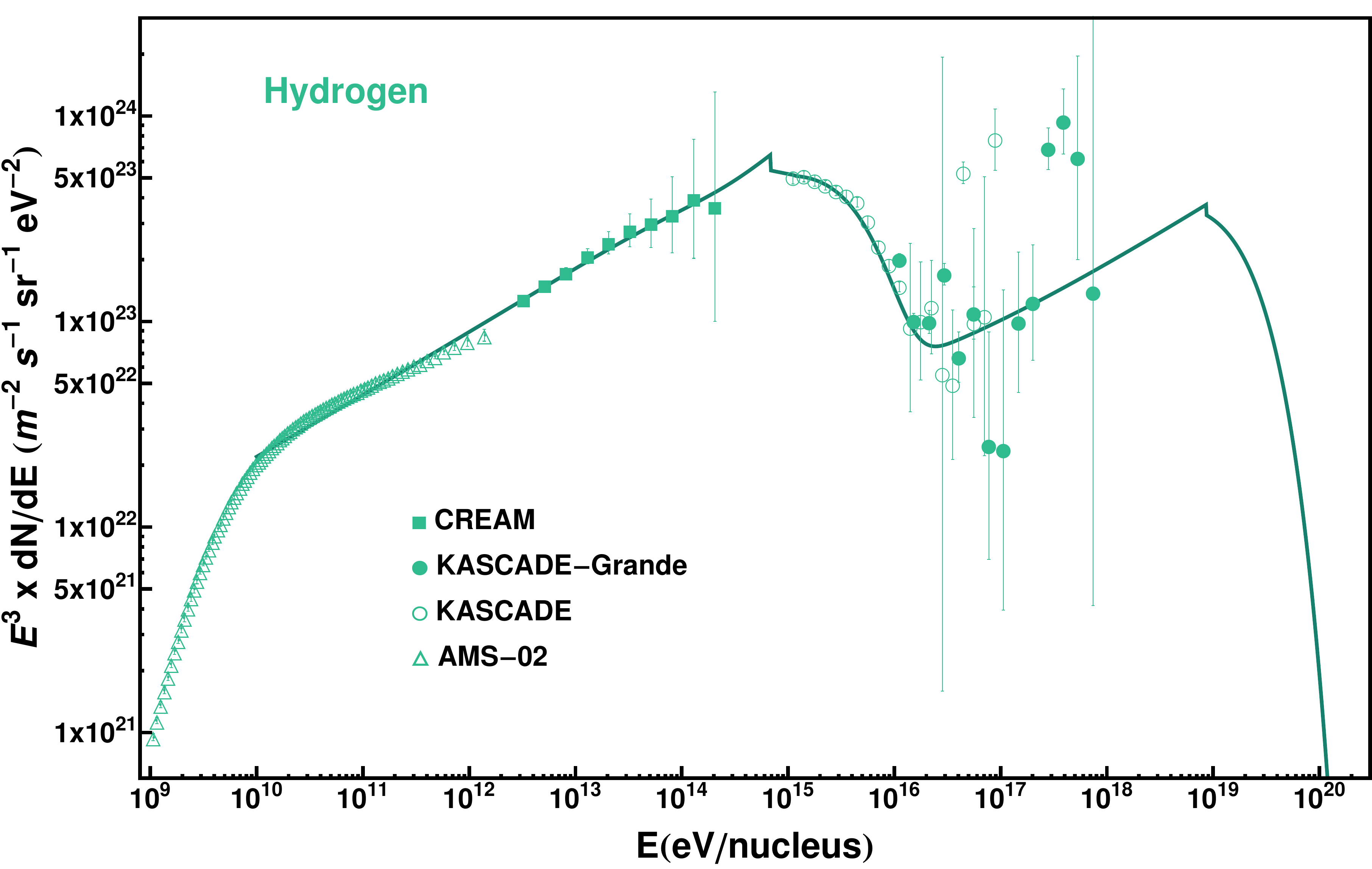}}
\caption{Hydrogen nuclei flux as a function of energy. The CREAM~\cite{CREAM_2041-8205-714-1-L89, CREAM_0004-637X-707-1-593}, AMS-02~\cite{AMS-02_ICRC_2013}, KASCADE~\cite{bib:kascade:pr:knee:1,bib:kascade:pr:knee:2} and KASCADE-Grande~\cite{bib:kascade:grande:thesis, bib:kascade:grande:icrc2011} data are shown together with the prediction of the model considered here. The model was fit to the data and the extrapolation to the highest energies was done following reference~\cite{Biermann_Vitor_2012}}
\label{fig:spectrum:h}
\end{figure}

\begin{figure}
\centerline{\includegraphics[width=13cm]{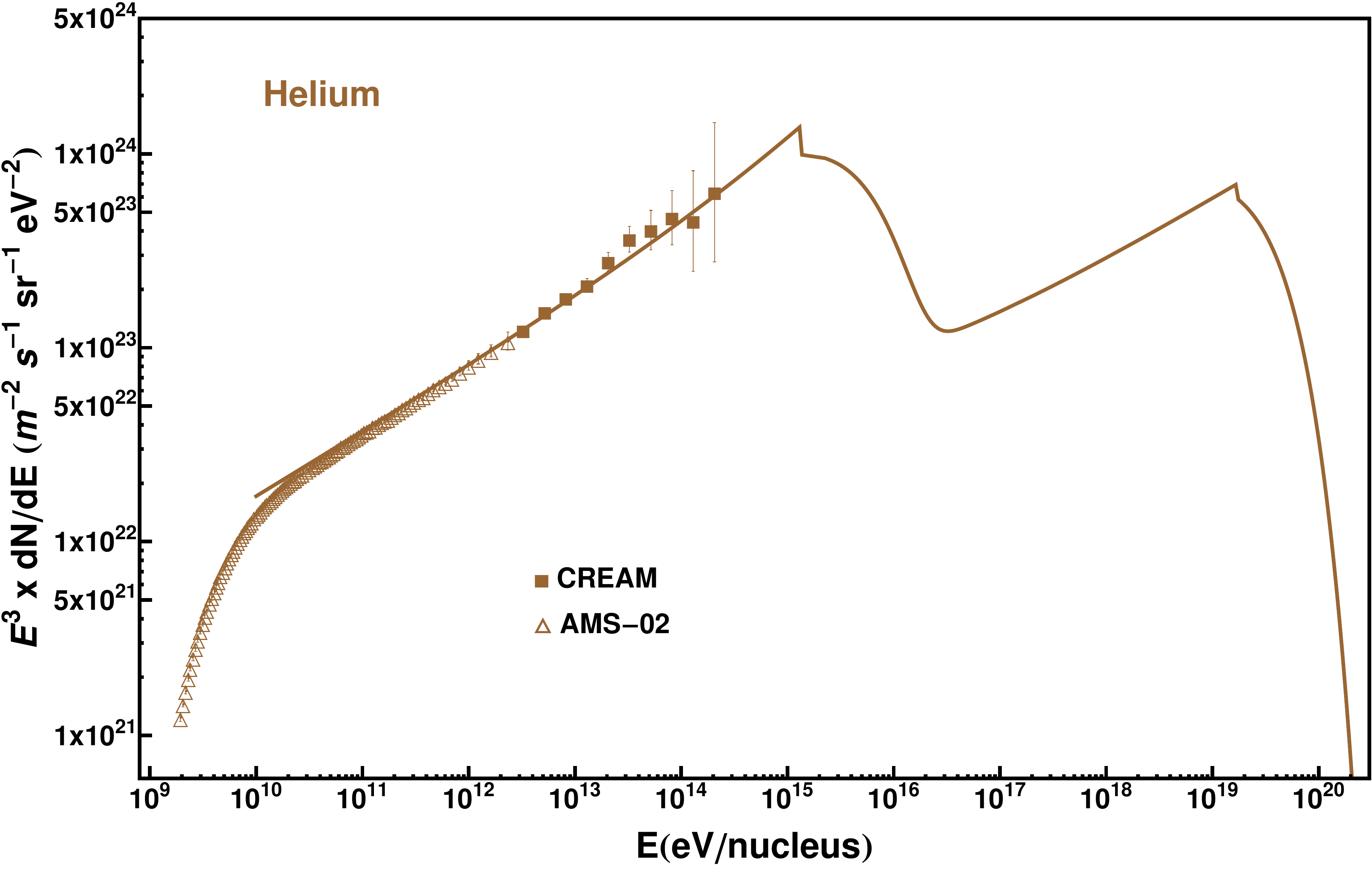}}
\caption{Helium nuclei flux as a function of energy. The CREAM~\cite{CREAM_2041-8205-714-1-L89, CREAM_0004-637X-707-1-593} and AMS-02~\cite{AMS-02_ICRC_2013} data are shown together with the prediction of the model considered here. The model was fit to the data and the extrapolation to the highest energies was done following reference~\cite{Biermann_Vitor_2012}. All energy breaks and cutoffs follow the rigidity dependency after the hydrogen fit shown in figure~\ref{fig:spectrum:h}.}
\label{fig:spectrum:he}
\end{figure}

Data from AMS-02~\cite{AMS-02_ICRC_2013} and CREAM~\cite{CREAM_2041-8205-714-1-L89, CREAM_0004-637X-707-1-593} have been used to validate the model in the energy range from \energy{10} $<E<$ \energyEV{15}. Both experiments published the H and He flux as shown in~\cref{fig:spectrum:h,fig:spectrum:he}. The data from both experiments offers a very hard constraint to the normalization constants ($A_1$) for H ($A_1^H$) and He  ($A_1^{He}$) elements. Once these parameters are set to the AMS-02 and CREAM data the relative contribution of these elements to the total flux are kept constant in the entire energy range studied in this paper (\energy{10} $<E<$ \energyEV{20}). Propagation effects, i.e., photo-nuclear disintegration might change the relative contribution of each element arriving on Earth at the highest energies $E>10^{17}$ eV.

The energy cutoff step at $E_2^{H-break}$ cannot be determined due to the lack of data in the energy range between \energy{14} and \energyEV{15}. However, the data measured by KASCADE~\cite{bib:kascade:pr:knee:1,bib:kascade:pr:knee:2} with energy above \energyEV{15} can be seen together with the AMS-02 and CREAM data for H in figure~\ref{fig:spectrum:h}. The figure the agreement in the relative flux of H as measured by the three experiments is clear from the figure. The agreement of the model to the data is also remarkable. The data from KASCADE was used to calculate $E_2^{H-cutoff} = 1.96 \times 10^{15}$ eV which is the energy of the knee of Hydrogen. Using the rigidity dependence of the model $E_2^{cutoff-e} = Z \times E_2^{H-cutoff}$ the energy breaks of other elements are determined. Figure~\ref{fig:spectrum:h} also shows the data of the KASCADE-Grande experiment~\cite{bib:kascade:grande:thesis, bib:kascade:grande:icrc2011}. The model is able to describe the connection between the KASCADE and KASCADE-Grande data which shows a continuous reduction of the flux up to $3\times 10^{16}$ eV. Beyond this energy, the KASCADE-Grande data suggest an flattening of the proton flux~\cite{bib:kascade:grande:ankle}. This energy sets the change of predominance from component 2 to component 4 as show in figure~\ref{fig:theory:original}. Again the rigidity dependency model is used to set the energy beyond which the extragalactic flux is predominant.

\begin{figure}
\centerline{\includegraphics[width=13cm]{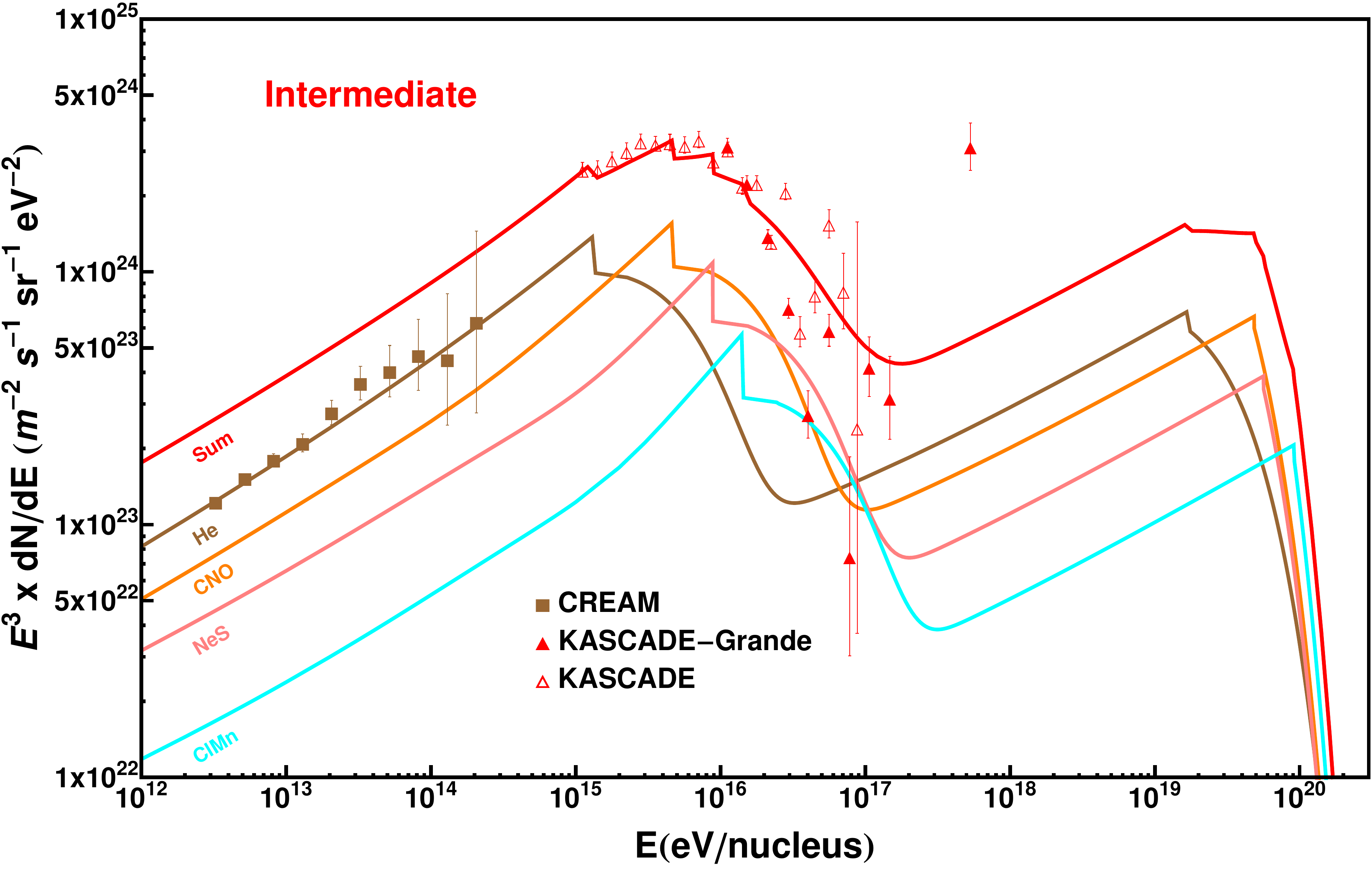}}
\caption{Intermediate nuclei (He, CNO, NeS and ClMn) flux as a function of energy. The CREAM~\cite{CREAM_2041-8205-714-1-L89, CREAM_0004-637X-707-1-593}, KASCADE~\cite{bib:kascade:pr:knee:1,bib:kascade:pr:knee:2} and KASCADE-Grande~\cite{bib:kascade:grande:thesis, bib:kascade:grande:icrc2011} data is shown together with the prediction of the model considered here. The model was fit to the data and the extrapolation to the highest energies was done following reference~\cite{Biermann_Vitor_2012}. All energy breaks and cutoffs follow the rigidity dependency after the hydrogen fit shown in figure~\ref{fig:spectrum:h}.}
\label{fig:spectrum:intermediate}
\end{figure}

The rigidity dependency and the relative flux of CNO, NeS and ClMn can be verified and adjusted using the KASCADE and KASCADE-Grande data~\cite{bib:kascade:grande:thesis}. Figure~\ref{fig:spectrum:intermediate} shows the energy spectrum of intermediate mass particles as measured by KASCADE and KASCADE-Grande experiments. The lines shown for CNO, NeS and ClMn are the result of the fit of the original model to the data. Since the energy breaks have been set by fitting the H spectrum and using the rigidity model dependency, the only free parameters in the fit are the normalization of each element flux. The agreement of the model to the data is very good. The spikes in the flux of each element caused by $E_2^{break}$ are visible in the sum of all intermediate elements with small amplitude. Unfortunately the resolution of the measurement is not enough to test the small spikes.

\begin{figure}
\centerline{\includegraphics[width=13cm]{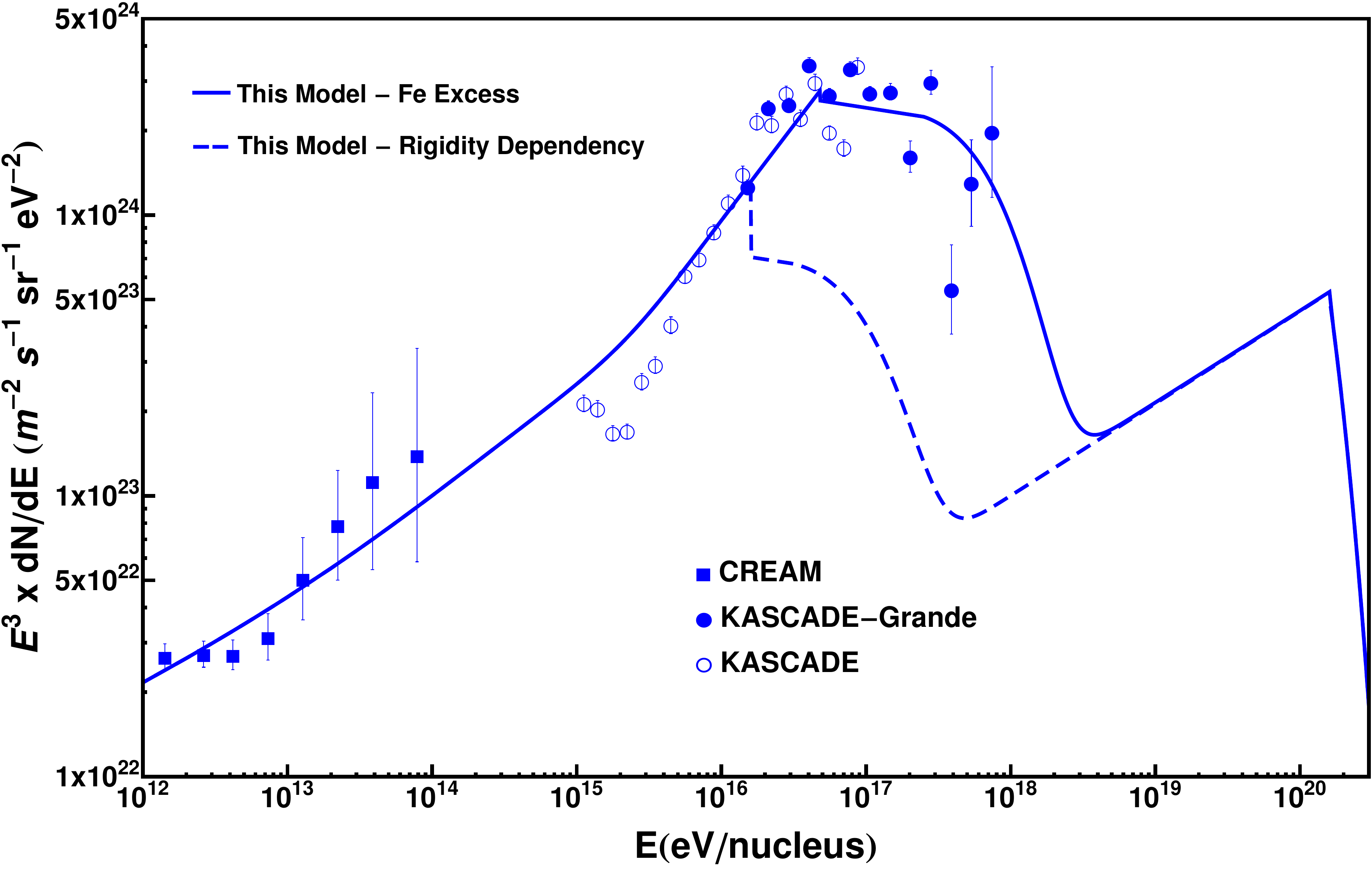}}
\caption{Iron nuclei flux as a function of energy. The CREAM~\cite{CREAM_2041-8205-714-1-L89, CREAM_0004-637X-707-1-593}, KASCADE~\cite{bib:kascade:pr:knee:1,bib:kascade:pr:knee:2} and KASCADE-Grande~\cite{bib:kascade:grande:thesis, bib:kascade:grande:icrc2011} data are shown together with the two hypotheses based on the model considered here. ``This model - Rigidity Dependency'' is used to identify the hypothesis in which $E_2^{cutoff-e} = Z \times E_2^{cutoff-H}$ for all elements. ``This model - Fe excess'' is used to identify the hypothesis in which $E_2^{cutoff-e} = Z \times E_2^{cutoff-H}$ for all elements except Fe. The model was fit to the data and the extrapolation to the highest energies was done following reference~\cite{Biermann_Vitor_2012}.}
\label{fig:spectrum:iron}
\end{figure}

Finally the model was compared to the iron flux measured by CREAM, KASCADE and KASCADE-Grande. Figure~\ref{fig:spectrum:iron} shows the fit of the model to the data considering two approaches. First the rigidity dependency of the energy cutoff ($E_2^{cutoff}$) was kept ($E_2^{cutoff-Fe} = 26 \times E_2^{cutoff-H}$) as shown by the dashed line. The label ``This model - Rigidity Dependency'' is used to identify the hypothesis in which $E_2^{cutoff-e} = Z \times E_2^{cutoff-H}$ for all elements. It is clear that in this case the model does not describe the KASCADE and KASCADE-Grande data. If the rigidity dependency is not kept and $E_2^{cutoff-Fe}$ is allowed to vary, the model describes fairly well the data as shown by the full line in figure~\ref{fig:spectrum:iron}. In this case the fit leads to $E_2^{cutoff-Fe} = 2.54 \times 10^{17}$ eV. The label ``This model - Fe excess'' is used to identify the hypothesis in which $E_2^{cutoff-e} = Z \times E_2^{cutoff-H}$ for all elements except Fe.

This analysis illustrates two possibilities to this model and at some extent to any rigidity dependent model conceived to describe the energy spectrum of cosmic rays with energy between \energy{15} and \energyEV{18}. The energy spectrum of iron nuclei measured by KASCADE and KASCADE-Grande seems to require an extra flux of heavy particles for energies between \energy{17} and \energyEV{18}. This extra flux can be provided if the rigidity dependency of the knees is not maintained or if an extra flux of iron from a yet unknown source is produced.

\begin{figure}
\centerline{\includegraphics[width=13cm]{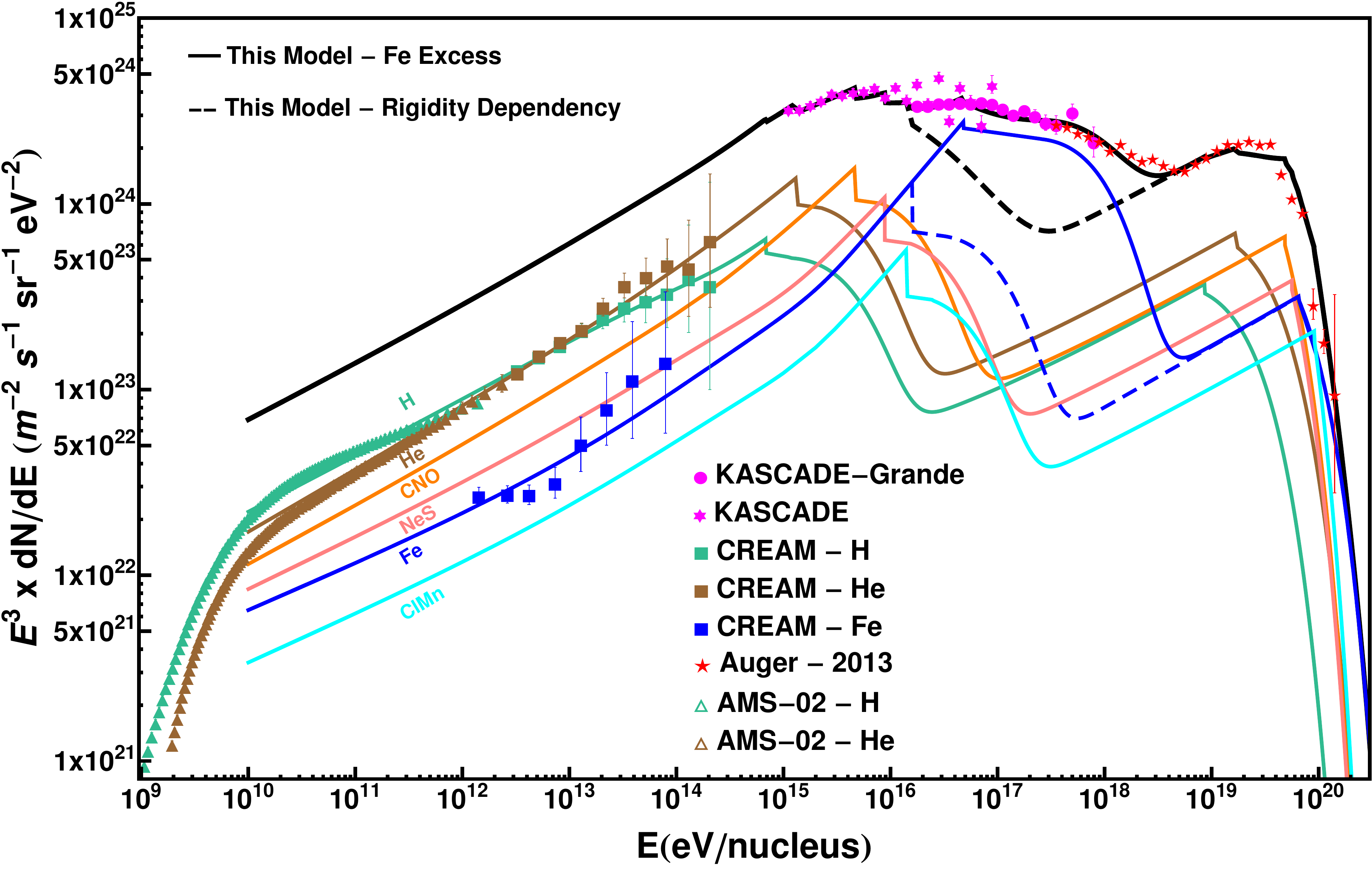}}
\caption{All particle flux as a function of energy. The CREAM~\cite{CREAM_2041-8205-714-1-L89, CREAM_0004-637X-707-1-593}, AMS-02\cite{AMS-02_ICRC_2013}, KASCADE~\cite{bib:kascade:pr:knee:1,bib:kascade:pr:knee:2} and KASCADE-Grande~\cite{bib:kascade:grande:thesis, bib:kascade:grande:icrc2011} and Auger~\cite{bib:auger:spectrum} data is shown together with the prediction of the model considered here. The model was fit to the data and the extrapolation to the highest energies was done following reference~\cite{Biermann_Vitor_2012}.``This model - Rigidity Dependency'' is used to identify the hypothesis in which $E_2^{cutoff-e} = Z \times E_2^{cutoff-H}$ for all elements. ``This model - Fe excess'' is used to identify the hypothesis in which $E_2^{cutoff-e} = Z \times E_2^{cutoff-H}$ for all elements except Fe. The model was fit to the data and the extrapolation to the highest energies was done following reference~\cite{Biermann_Vitor_2012}.}
\label{fig:spectrum:total}
\end{figure}

The prediction of the model as fitted to the energy spectra of the elements shown in figure~\cref{fig:spectrum:h,fig:spectrum:intermediate,fig:spectrum:iron} was summed in order to obtain the total flux of particles. Figure~\ref{fig:spectrum:total} shows the energy spectrum of all particles as measured by KASCADE~\cite{bib:kascade:pr:knee:1}, KASCADE-Grande~\cite{bib:kascade:grande:ankle} and The Pierre Auger Observatory~\cite{bib:auger:spectrum}. The spectra of each element as measured by CREAM (H, He and Fe) and AMS-02 (H and He) are also shown. Two possibilities for the total flux predicted by the model are shown. The full black line takes into account the iron flux that fits the KASCADE-Grande data better, which does not obey the rigidity dependency of the knee. The dashed black line takes into account the iron flux that fits the KASCADE-Grande data worse which retained the rigidity dependency of the knee.

\section{Comparison of the model to the depth of shower maximum}
\label{sec:mass}

\begin{figure}
\centerline{\includegraphics[width=13cm]{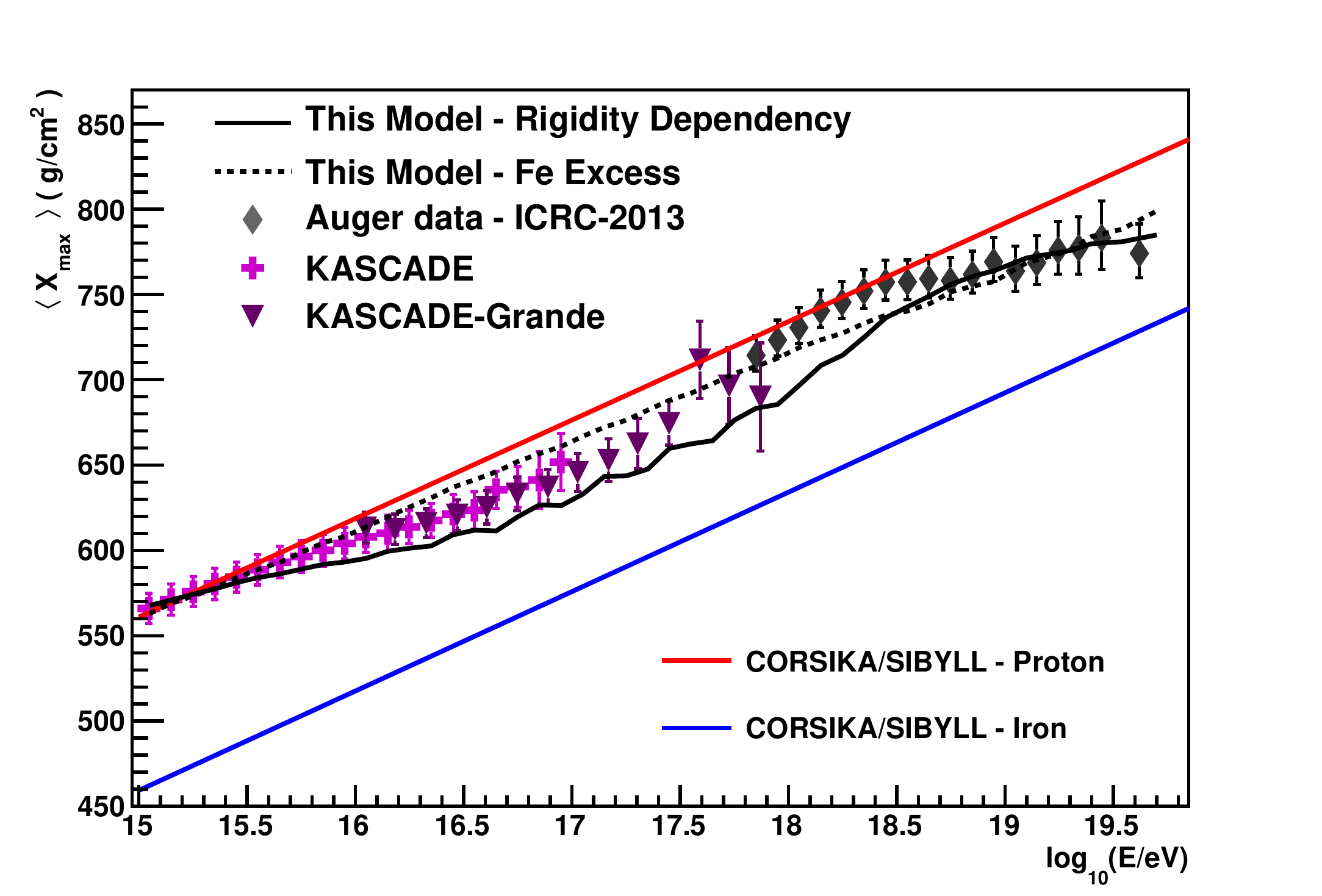}}
\caption{Mean depth of shower maximum (\meanXmax) as a function of energy. KASCADE~\cite{bib:kascade:pr:knee:1,bib:kascade:pr:knee:2} and KASCADE-Grande~\cite{bib:kascade:grande:thesis, bib:kascade:grande:icrc2011} data were converted to \meanXmax using reference~\cite{ToderoPeixoto201318}. The Auger data are also shown~\cite{bib:auger:xmax}.``This model - Rigidity Dependency'' is used to identify the hypothesis in which $E_2^{cutoff-e} = Z \times E_2^{cutoff-H}$ for all elements. ``This model - Fe excess'' is used to identify the hypothesis in which $E_2^{cutoff-e} = Z \times E_2^{cutoff-H}$ for all elements except Fe.}
\label{fig:xmax}
\end{figure}

For energies above \energyEV{18}, the most reliable composition parameter is the depth of the shower maximum (\xmax). The results published by the Pierre Auger Collaboration~\cite{bib:auger:xmax} concerning the evolution of the \xmax with energy is independent of shower simulation and detector efficiencies, therefore this datum is used for comparison to the model predictions. The KASCADE and KASCADE-Grande data can also be transformed to an equivalent \meanXmax. Using the parametrization of \meanXmax as a function of energy and mass published in reference~\cite{ToderoPeixoto201318} and the flux of element groups measured by the experiments~\cite{bib:kascade:pr:knee:1,bib:kascade:pr:knee:2,bib:kascade:grande:prl} it is possible to calculate the equivalent \meanXmax of the measured abundance. Figure~\ref{fig:xmax} shows the comparison of the model as optimized in the previous section to the transformed KASCADE data, transformed KASCADE-Grande data, and Auger data. The energy spectra predicted by the model was transformed into \meanXmax measurements using the same procedure adopted for the KASCADE and KASCADE-Grande data, using the parametrization of \meanXmax as a function of energy and mass published in reference~\cite{ToderoPeixoto201318} for the Sibyll hadronic interaction model~\cite{bib:sibyll}.

Both versions of the model with and without an extra iron component describe the data qualitatively well. The Fe excess artificially introduced around \energyEV{17} produces a decrease of the \meanXmax and makes the transition between the transformed KASCADE-Grande data and Auger data more abrupt.

\section{Final Remarks}
\label{sec:conclusion}

The data analyzed here confirm the validity of the model in which only the Galaxy and CenA could produce all cosmic rays measure on Earth with energy above \energyEV{10}. For the first time, the data from AMS-02 and CREAM were used to fix the relative contribution of individual elements.

The analysis presented here tries to describe the iron nuclei flux reconstructed by the KASCADE-Grande experiment within the framework of the original model~\cite{Stanev_Biermann_Gaisser_1993}. The calculations shown in the previous section suggest an extra flux of iron nuclei which does not obey the rigidity dependence hypothesis. The KASCADE-Grande Collaboration reported a suppression in the flux of heavy elements at $8\times 10^{16}$ eV~\cite{bib:kascade:grande:prl}. However, according to this publication the suppression is less significant in the all particle spectrum~\cite{bib:kascade:grande:prl}. At the same time, it is clear that the steepening of the spectrum after the knee is more severe for the light component rather than for the heavy component. The calculation done here suggests that this energy range $10^{16.5} < E < 10^{17.5}$ eV might contain an extra flux of a heavy element. This idea has been advocated by Hillas~\cite{bib:hillas}. However, in his proposal, Galactic magnetars are considered as possible sources for the extra iron flux. Another similar study was presented in reference~\cite{1475-7516-2014-10-020} in which the need of an additional extragalactic component is pointed out.

Depending on how many supernova explosions contribute cosmic ray particles near the knee region and beyond, the possibility cannot be excluded with certainty, that just relatively few explosions contribute, and then one is the strongest. If this strongest one derives from a more powerful supernova, such as in the hyper-nova model discussed in reference~\cite{bib:paczynski,bib:wang} then the cosmic ray spectrum near and beyond the knee region could be modified.  This could entail, that the corresponding knee energy, the spectral index below the knee, the polar cap component, and the component beyond the knee could all be changed, especially by being shifted to higher energies, and that the deeper layers of the stars might be exposed, allowing Fe to become much stronger for this one star.

The reconstructed particle fluxes published by KASCADE and KASCADE-Grande were converted to an equivalent \meanXmax as shown together with the measured \meanXmax published by the Pierre Auger Collaboration. Despite the large error bars of the converted KASCADE-Grande data for energies above \energy{17.5},the apparent continuity of the KASCADE-Grande to Auger data is remarkable. The hypothesis with and without an extra iron nuclei flux for the model proposed here are compared to the data. The data cannot to discriminate between the two hypothesis. Here we see a possible trade-off, a special Fe-component, well supported by data, and a normal Fe-component (i.e. as in the rigidity dependent model), and a much enhanced high energy H-component. New composition data from the LOFAR, TA and Auger experiments is expected to the published in the near future for the energy range \energy{17} to \energyEV{18}. These new data might be able to discriminate between the two hypotheses.

\section{Acknowledgments}

CJTP thanks the support of CENAPAD computer center, FAPESP and CNPq fellowships. VdS thanks the support of the Brazilian population via CNPq and FAPESP (2012/22540-4). PLB acknowledges discussions with J. Becker, J. Bl{\"u}mer, L. Caramete, R. Engel, J. Everett, H. Falcke, T.K. Gaisser, L.A. Gergely, A. Haungs, S. Jiraskova, H. Kang, K.-H. Kampert, A. Kogut, Gopal Krishna, R. Lovelace,  K. Mannheim, I. Maris, G. Medina-Tanco, A. Meli, B. Nath, A. Obermeier, J. Rachen, M. Romanova, D. Ryu, E.-S. Seo, T. Stanev, P. Wiita, W.R. Binns, T. Frederico, and J. Var. The authors acknowledge their KASCADE, KASCADE-Grande and Pierre Auger Collaborators and the AMS-02 and CREAM experiments. The authors thank the Pierre Auger Collaboration for permission to use their data prior to journal publication.

\bibliographystyle{unsrt}\small
\bibliography{allbib.bib}


\newpage

\end{document}